\documentclass[
    ,final            
  ]
  {aipproc}

\usepackage{amssymb,amsmath}
\layoutstyle{8x11double}
\usepackage[T1]{fontenc}

\def\gam{{VHE \mbox{$\gamma$-ray} }}
\def\rays{{VHE \mbox{$\gamma$-rays} }}
\def\sgr{{\mbox{Sgr A$^{*}$}}}

\begin{document}

\title{Time-dependent absorption of very high-energy gamma-rays
  from the Galactic center by pair-production}

\classification{95.10.Eg, 95.30.Cq, 95.85.Pw}
\keywords      {Galactic center, S-stars, absorption by pair-production, $\gamma$-rays}

\author{Attila Abramowski}{
  address={Department of Physics, Institute for Experimental Physics,
    University of Hamburg, Luruper Chaussee 149, D-22761 Hamburg}
}

\author{Dieter Horns}{
  address={Department of Physics, Institute for Experimental Physics, University of Hamburg, Luruper Chaussee 149, D-22761 Hamburg}
}

\author{Stefan Gillessen}{
  address={MPI for Extraterrestrial Physics Garching, Giessenbachstrasse, D-85748 Garching}
}

\author{Joachim Ripken}{
  address={Department of Physics, Institute for Experimental Physics, University of Hamburg, Luruper Chaussee 149, D-22761 Hamburg}
}

\author{Christopher van Eldik}{
  address={MPI for Nuclear Physics Heidelberg, Saupfercheckweg 1, D-69117 Heidelberg}
}

\begin{abstract}
Very high energy (VHE) gamma-rays have been detected from the direction of the Galactic
center. The H.E.S.S. Cherenkov telescopes have located this $\gamma$-ray
source with a preliminary position uncertainty of 8.5" per axis (6"
statistic + 6" sytematic per axis). Within the uncertainty
region several possible counterpart candidates exist: the Super
Massive Black Hole \sgr, the Pulsar Wind Nebula candidate
G359.95-0.04, the Low Mass X-Ray Binary-system J174540.0-290031, the stellar cluster IRS
13, as well as self-annihilating dark matter. It is experimentally
very challenging to further improve the positional accuracy in this
energy range and therefore, it may not be possible to clearly associate one of the counterpart candidates with the
VHE-source. Here, we present a new method to investigate a possible link
of the VHE-source with the near environment of \sgr (within approximately
1000 Schwarzschild radii). This method uses the time- and
energy-dependent effect of absorption of \rays by pair-production (in
the following named pair-eclipse) with low-energy photons of stars closely orbiting the SMBH \sgr. 
\end{abstract}

\maketitle


\section{Introduction}

\begin{figure}
  \includegraphics[height=.3\textheight]{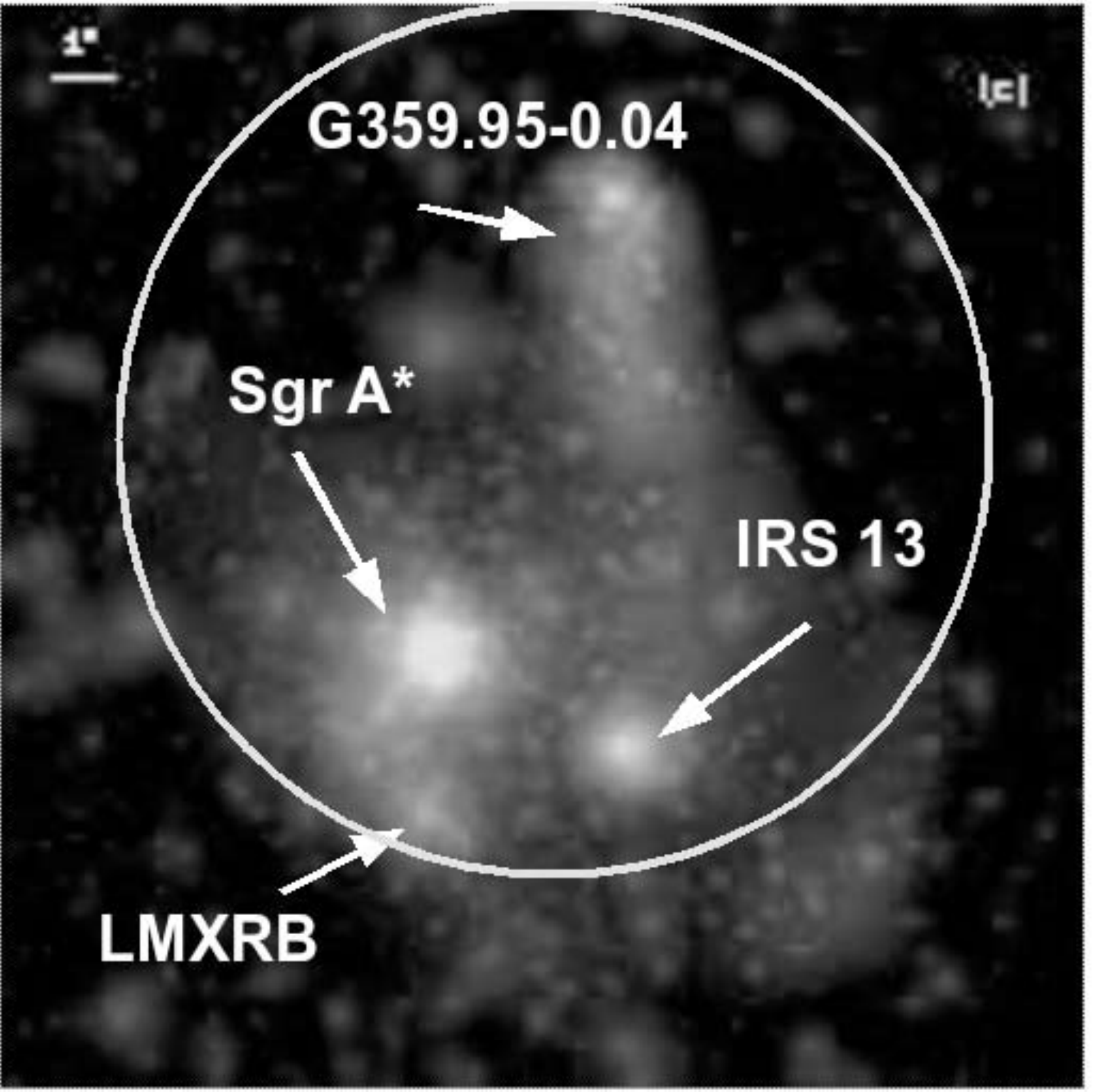}
  \caption{Chandra X-ray and NACO Near-infrared Composite of the
    Galactic center region \cite{2006MNRAS.367..937W}. The circle
    marks the systematic (and therefore irreducable) H.E.S.S. uncertainty region of $6"$.}
  \label{fig:chandra}
\end{figure}

Because of its vicinity of about 8 kpc, the center of our own galaxy
offers a unique possibility to study a galactic nucleus. Especially the
super massive black hole (SMBH) in the center (for a recent review see
e.g. \cite{2008arXiv0808.2624R}) and the objects in its direct neighborhood can be
observed with high precision compared to distant galaxies. While the
observation of the Galactic center (GC) in the optical waveband
suffers strongly from obscuration by dust, it has been studied
intensely at radio, infrared,
X-ray and $\gamma$-ray wavelengths.
The H.E.S.S. telescopes have
detected \rays from the direction of the GC in an energy range between
165 GeV and 10 TeV\cite{2006Natur.439..695A}. One of the two sources reported there,
\mbox{HESS J1745-290}, is coincident with Sgr A* within the positional
errors. Despite the unprecedented accuracy achieved, the
identification with a counterpart observed at other wavelengths is difficult
because the observation has a systematic positional uncertainty of 6" \cite{2007arXiv0709.3729V}. In this region four possible counterparts
exist (see figure \ref{fig:chandra}). Additionally, the \rays could also originate
from self-annihilating dark matter. Even with the next
generations of Imaging air Cherenkov telescopes (IACT),
\mbox{H.E.S.S. II}, MAGIC 2 or the planned CTA, with a great
improvement of sensitivity, the positional accuracy (limited by
systematics) will remain the
same achieved today. So it stays
difficult to directly identify the \gam source.

Here, we present an indirect 
method which uses the \textit{timing} signature of absorption of \rays
by pair-production with low-energy photons of stars orbiting the
SMBH. This method is uniquely applicable to the GC because only here stars close to a SMBH are known. With this
method we can investigate a possible alignment of
the \rays with one of the candidates, the SMBH \sgr. In the first part
of this paper we present the orbits of the S-stars followed by a short review of the
pair-production process. Finally simulated lightcurves for present and
future IACTs will be presented.

\section{Stars orbiting the SMBH \sgr}

\begin{figure}
  \includegraphics[height=.21\textheight]{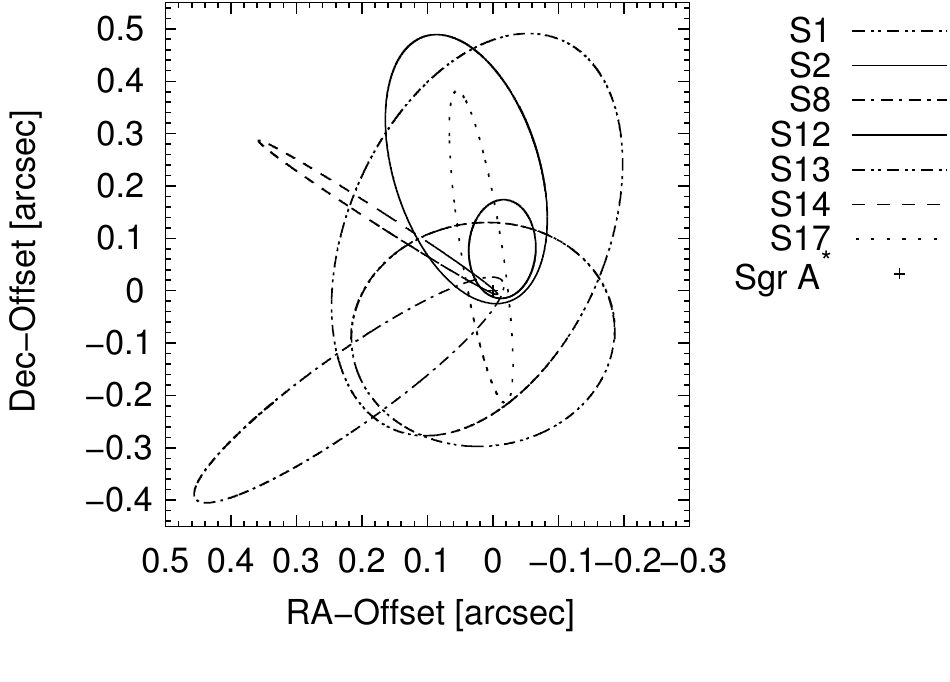}
  \caption{Projection on the sky of 7 S-Stars. ($0.1 \, \text{arcsec}$
    correspond to $10^{14} \,\text{m} \approx 10^{4}
    R_{Schwarz}$ at an assumed distance to GC of $7.62 \, \text{kPc}$)}
  \label{fig:orbits}
\end{figure}

\begin{figure}
  \includegraphics[height=.4\linewidth]{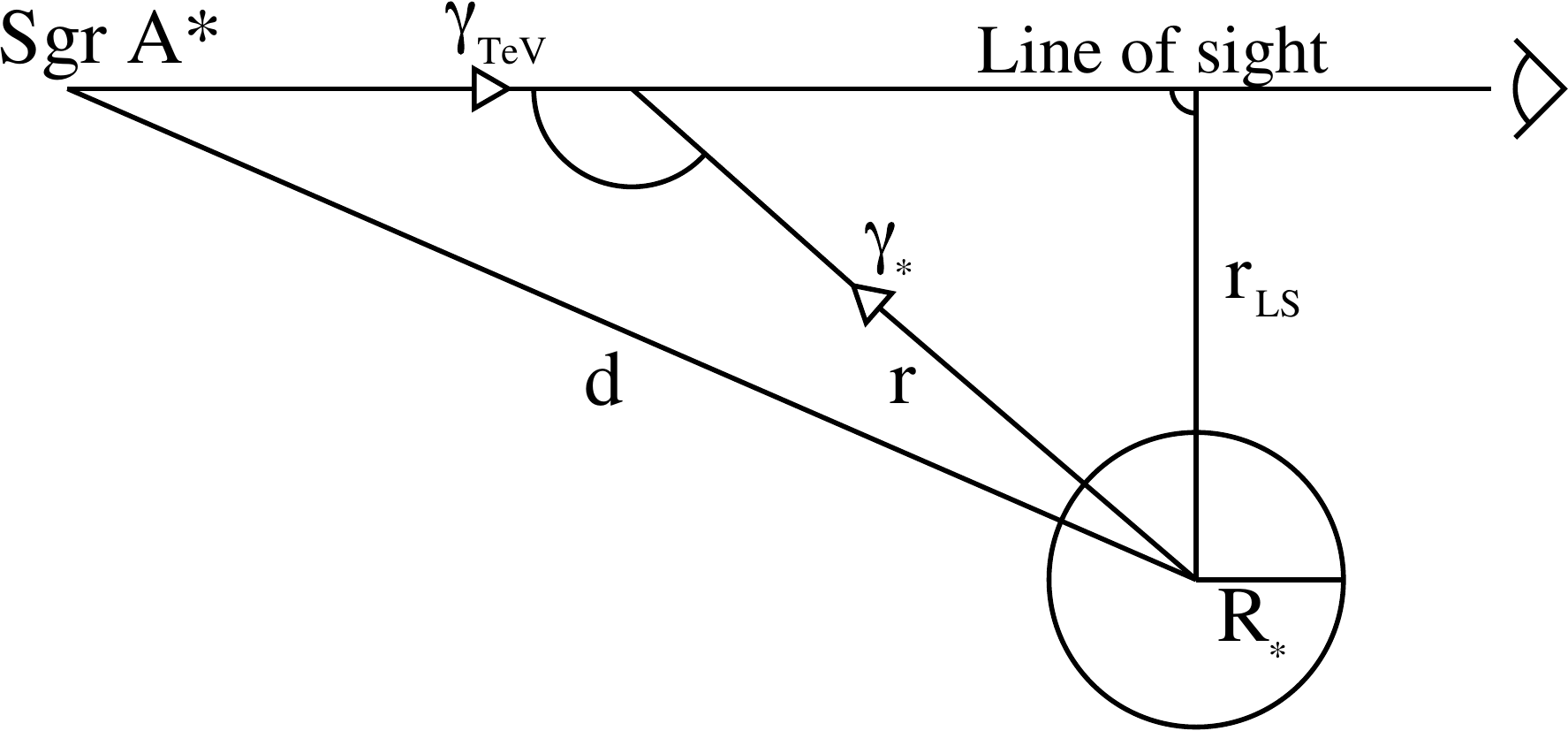}
  \caption{Schematic view of pair-production at the GC}
  \label{fig:scheme}
\end{figure}

From observations in the near-infrared 27 stars (so called S-Stars)
with a magnitude $K \lesssim 16$ are known to move on stable
Keplerian orbits very close ($<
0.7 \, \text{arcsec} \ \widehat{=} \ 8 \times 10^{14} \, \text{m}$)  to
\sgr. Using high
precision measurements with the VLT \cite{Eisenhauer2005} and the Keck
telescope \cite{2005ApJ...620..744G} the
orbital parameters and the stellar classes \cite{2008ApJ...672L.119M}
of some of 
these stars were determined. Figure \ref{fig:orbits} shows the
projected orbits of 7 S-stars calculated from
\cite{Gillessen2008}. Beside S17, which is a late-type star, these stars are identified as early type main-sequence stars of type B. For the calculations given here,
it is assumed that all stars produce a blackbody photon density $n$ at a
distance $r$ ($\varepsilon$ : photon energy)
\small
\begin{equation}
n(\varepsilon,r)=\frac{\varepsilon^{2}}{\hbar^{3}c^{3}\pi^{2}} \
\frac{1}{\exp(\varepsilon/kT_{*})-1} \  \frac{R^{2}_{*}}{4r^{2}} 
\label{eq:n}
\end{equation}
\normalsize
with a temperature of $T_{*}= 25000 \, \text{K}$ and a stellar radius of
$R_{*}=10.6 \, R_{\odot}$ equivalent to a
early type spectral class $\sim \text{B}1$ (being the mean value for
S2 \cite{2008ApJ...672L.119M}).  The factor
$R_{*}^{2}/4r^{2}$ is a dilution factor to take into account the
decrease with the distance $r$ (see figure \ref{fig:scheme}). Equation
\ref{eq:n} is of course only valid for $r
\gg R_{*}$ which is the case here ($R_{*}=7.37 \times 10^{9} \, \text{m}$ compared
to $r_{LS} > 10^{13} \, \text{m}$). 

\begin{table}
\begin{tabular}{lrrr}
\hline
  & \tablehead{1}{r}{b}{S2}
  & \tablehead{1}{r}{b}{S8}
  & \tablehead{1}{r}{b}{S14}\\
\hline
Orbital period [years] & $15.24\pm0.36$ & $67.2\pm5.5$ & $38.0\pm 5.7$\\
\hline
Minimal distance to \\ \sgr \  $d$ [mas] & $15.2$ & $24.2$  & $9.5$ \\
Time & $2017.377\pm0.012$ & $2053.92\pm0.81$ & $2045.222\pm0.002$\\
\hline
Minimal distance to \\ line of sight  $r_{LS}$ [mas] & $10.6$ & $16.1$ & $2.4$  \\
Maximal Absorption $e^{-\tau}$ & $0.993$ & $0.995$  & $0.974$ \\
Time & $2017.410$ & $2053.72$ & $2045.390$\\
\hline

\end{tabular}
\caption{Parameter of selected S-Stars}
\label{tab:stars}
\end{table}

Table \ref{tab:stars} shows some parameters of selected S-stars. The
values of the minimal distance $d$ and the minimal distance to the line of sight
$r_{LS}$ were derived from the orbital parameters. The estimated
uncertainties on the
distances are in the order of $10^{-3} \text{arcsec} =\text{mas}$.

\section{Absorption by pair-production}

\begin{figure}[b]
  \includegraphics[height=.23\textheight]{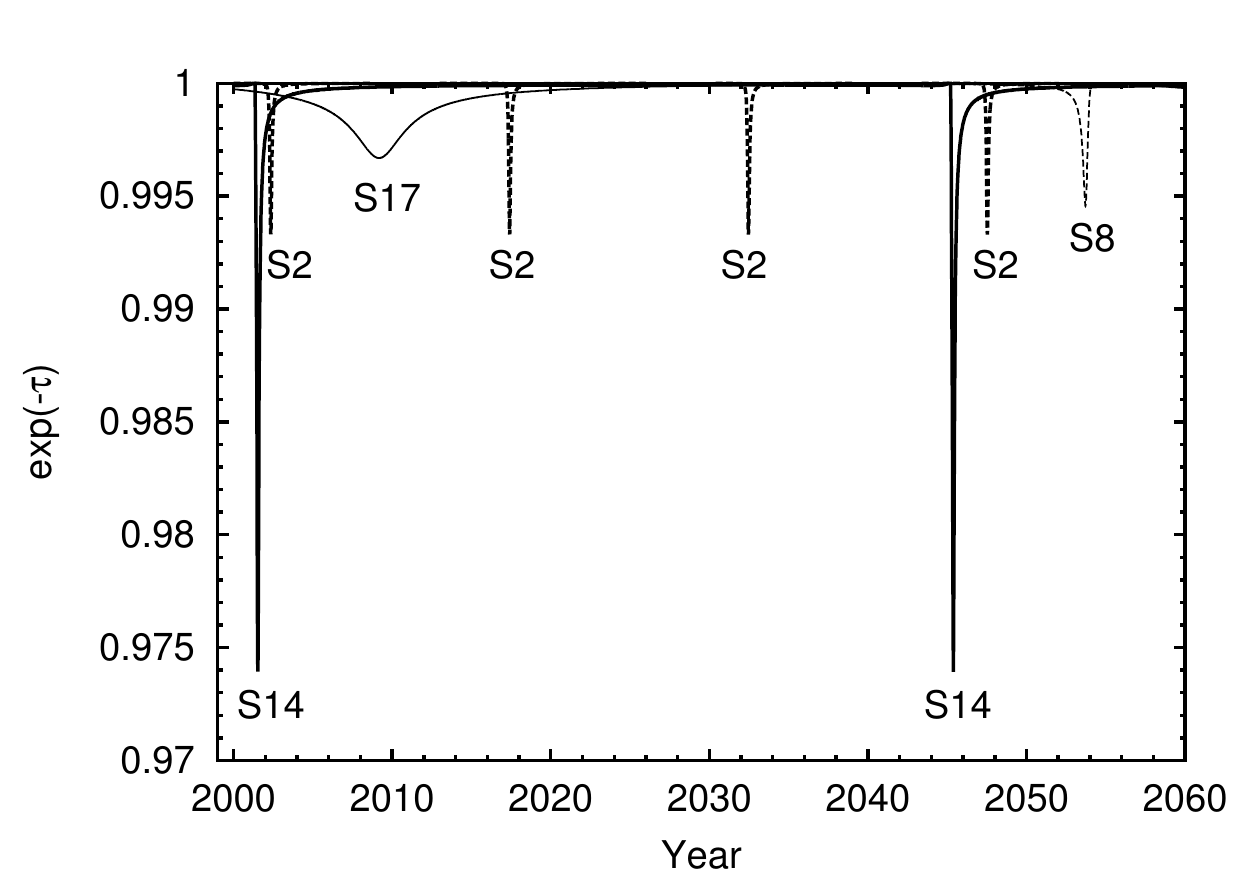}
  \caption{Predicted absorption as function of time (for maximal absorption
    effect at
    $E_{VHE}=200 \, \text{GeV}$)}
  \label{fig:abs}
\end{figure}

A schematic view of the underlying geometry assumed is shown in figure
\ref{fig:scheme}. A $\gamma$-ray propagating from \sgr \ interacts with a
low-energy photon from the S-star under an \mbox{angle $\theta$}. The relative position
of the line of sight (LoS) is defined by the distance
$r_{LS}$.

The process of pair-production ($\gamma+\gamma^{\prime}\rightarrow
e^{+}+e^{-}$) of two photons with energy
$\varepsilon$ and $E$ is well studied
\cite{1967PhRv..155.1404G}. The cross-section is given by equation
\eqref{eq:sigma} with an energy threshold of $\varepsilon_{min}=2m_{e}^{2}c^{4}/(E(1-\cos{\theta}))$.
\small

\begin{align}
 \label{eq:sigma}
\sigma(\varepsilon,E,\theta)&=\frac{3\sigma_T}{16}(1-\beta^2)\left[(3-\beta^4)\ln\frac{1+\beta}{1-\beta} \
2\beta(2-\beta^2)\right]\\
\beta&=\left( 1-\frac{2m_e^2c^4}{E\varepsilon(1-\cos(\theta))}\right) ^{1/2}\notag
\end{align}

\normalsize

In order to exactly calculate the optical depth $\tau$ for a $\gamma$-ray from
\sgr \ with the
low-energy photons from the S-star one has to integrate
over the line of sight, the surface of the star and the energy of the
low-energy photon. Here we use the point-source approximation for the star
emitting the low energy photons, reducing the calculation to a double
integral \cite{2006A&A...451....9D}:  
\small
\begin{equation}
\begin{split}
\tau(E,t)=&\int_{0}^{\theta_{\text{max}(t)}} d\theta  \ \left[-r(t) \
  \left(1+\frac{1}{\tan^{2}\theta}\right)\right] \\
&\int_{\varepsilon_{\text{min}}}^{\infty} d \varepsilon \
\sigma(\varepsilon,E,\theta) \ n (\varepsilon,r) \ (1- \cos
  \theta) 
\end{split}
\label{eq:tau}
\end{equation}
\normalsize

For $R_{S}\ll r_{LS}$ this is an appropriate approximation. Also the integration along the LoS is
changed to an integration over the \mbox{angle $\theta$} from 0 (photons are
parallel) to a maximal angle $\theta_{\text{max}}$ derived from the
distance from \sgr \   to the
star. The absorption is strongly energy dependent with a maximum at about
200 GeV for the assumed photon density. The distance $r(t)$ is derived
from the orbital parameters. The resulting optical depth $\tau$ is
time dependent because the photon density
$n$ changes with the star moving along the orbit. Figure \ref{fig:abs} shows the
result of the calculation for the S-stars. In Table \ref{tab:stars}
the values for the maximal absorption effect are shown. The error on
the absorption from the orbital uncertainties are under investigation. Additionally, the uncertainty on the spectral class of the star, as well as the relative
positional uncertainty of the radio position of \sgr ($\sim 2 \,
\text{mas}$ \cite{Gillessen2008}) will lead to an
even higher error for $\tau$. Only for the stars
S2, S8 and S14 the absorption effect is $>0.3\%$ and the largest
absorption is seen for S14 as expected ($\sim 3\%$).   

\begin{figure}
  \includegraphics[width=0.65\linewidth]{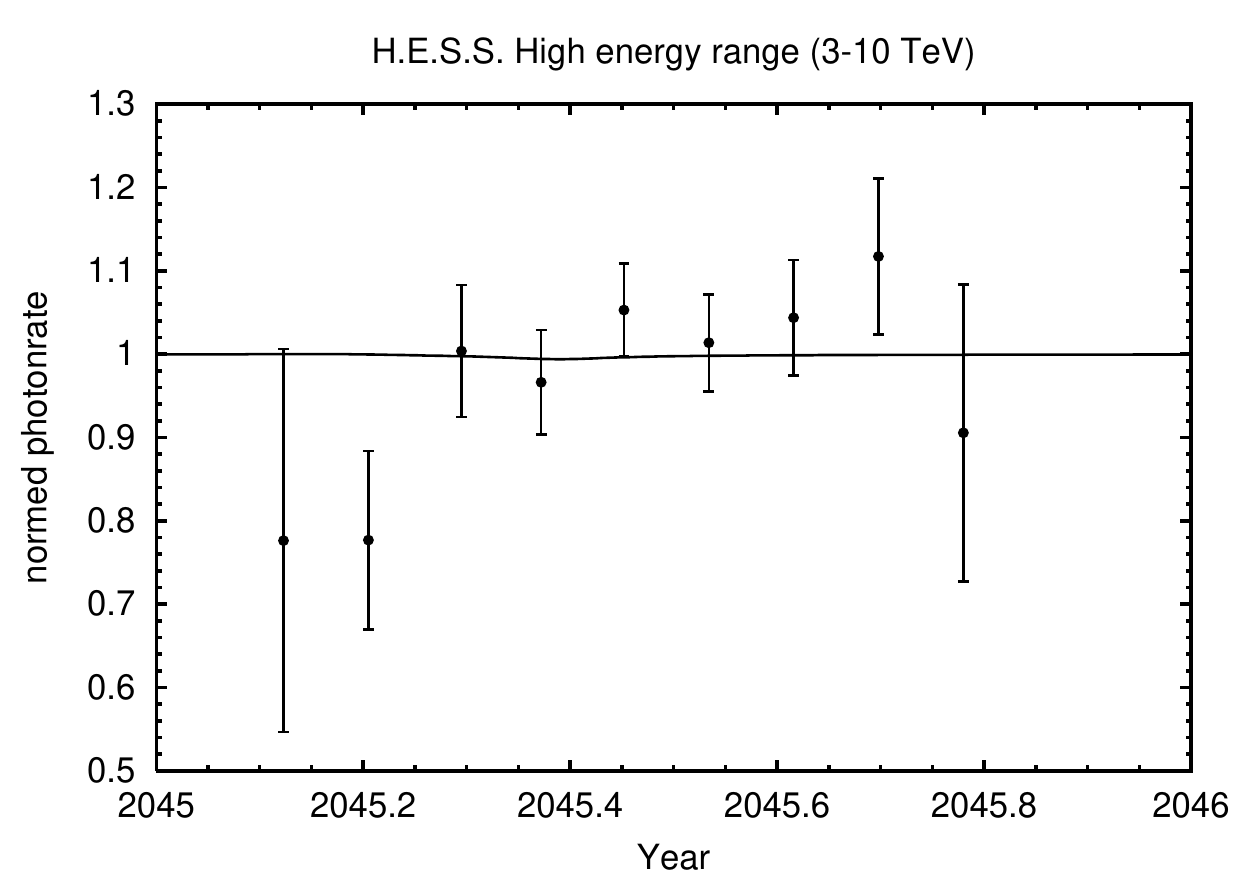} 
  \includegraphics[width=0.65\linewidth]{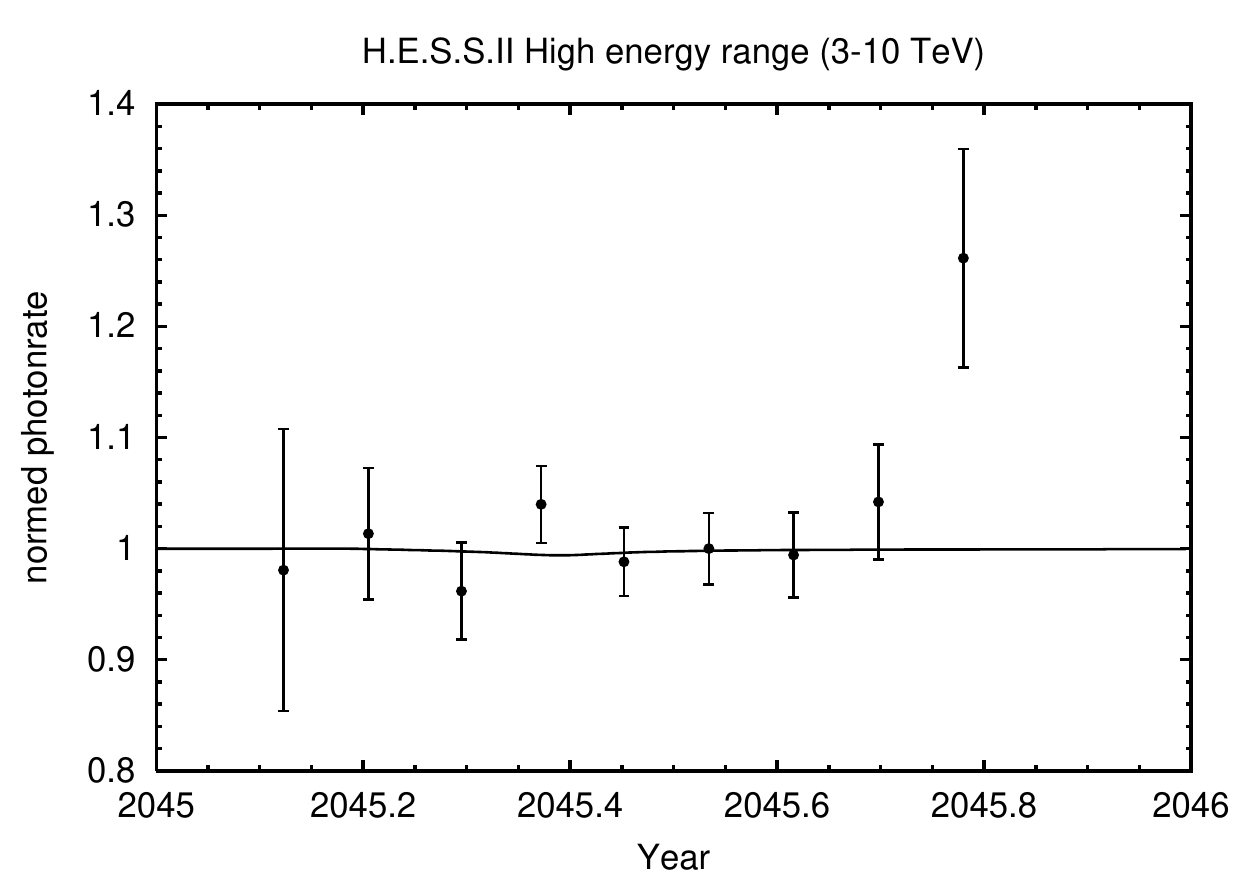} 
  \includegraphics[width=0.65\linewidth]{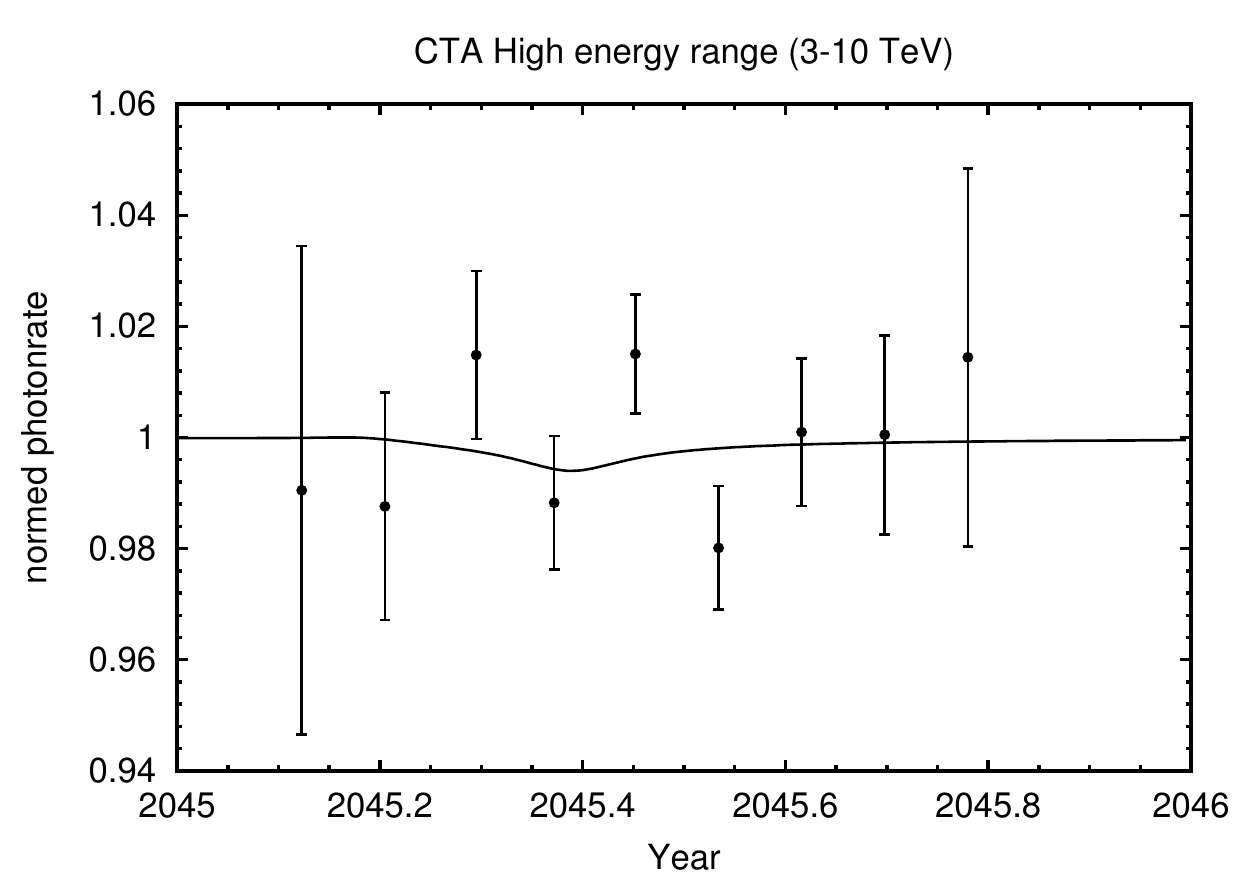} 
  \caption{Simulated Lightcurves: High-energy band 3-10 TeV. The
    solid line is the calculated absorption for 5 TeV.}
  \label{fig:lightcurve_high}
\end{figure}

\begin{figure}
  \includegraphics[width=0.65\linewidth]{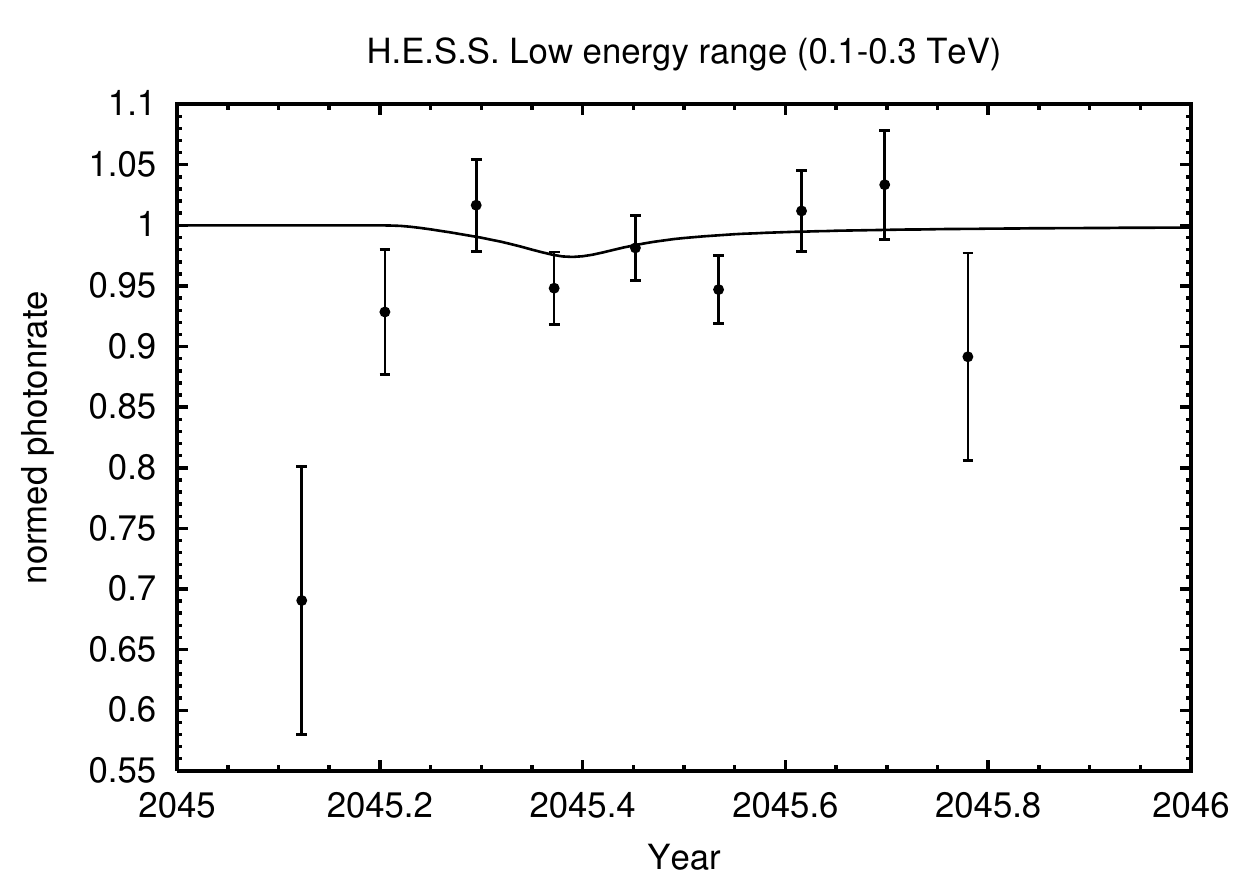} 
  \includegraphics[width=0.65\linewidth]{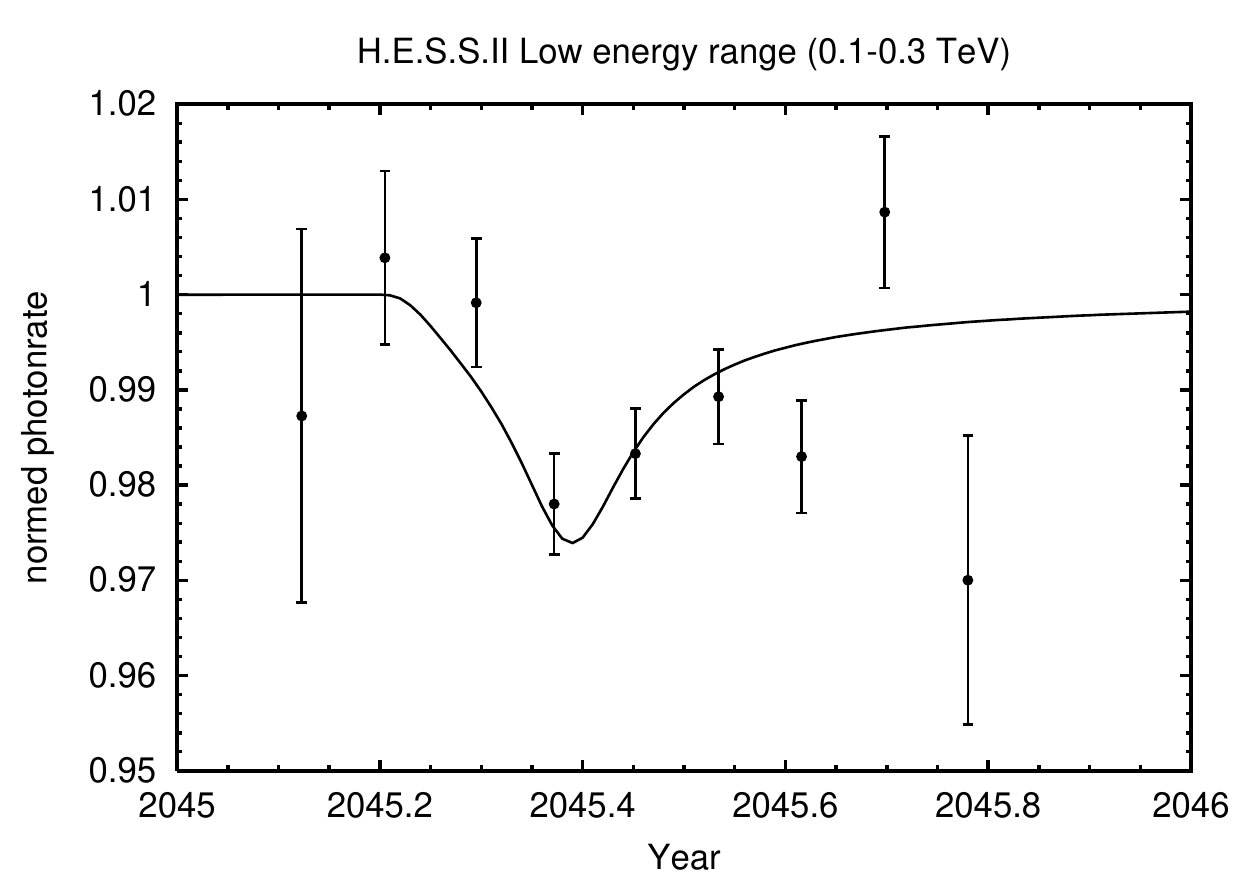} 
  \includegraphics[width=0.65\linewidth]{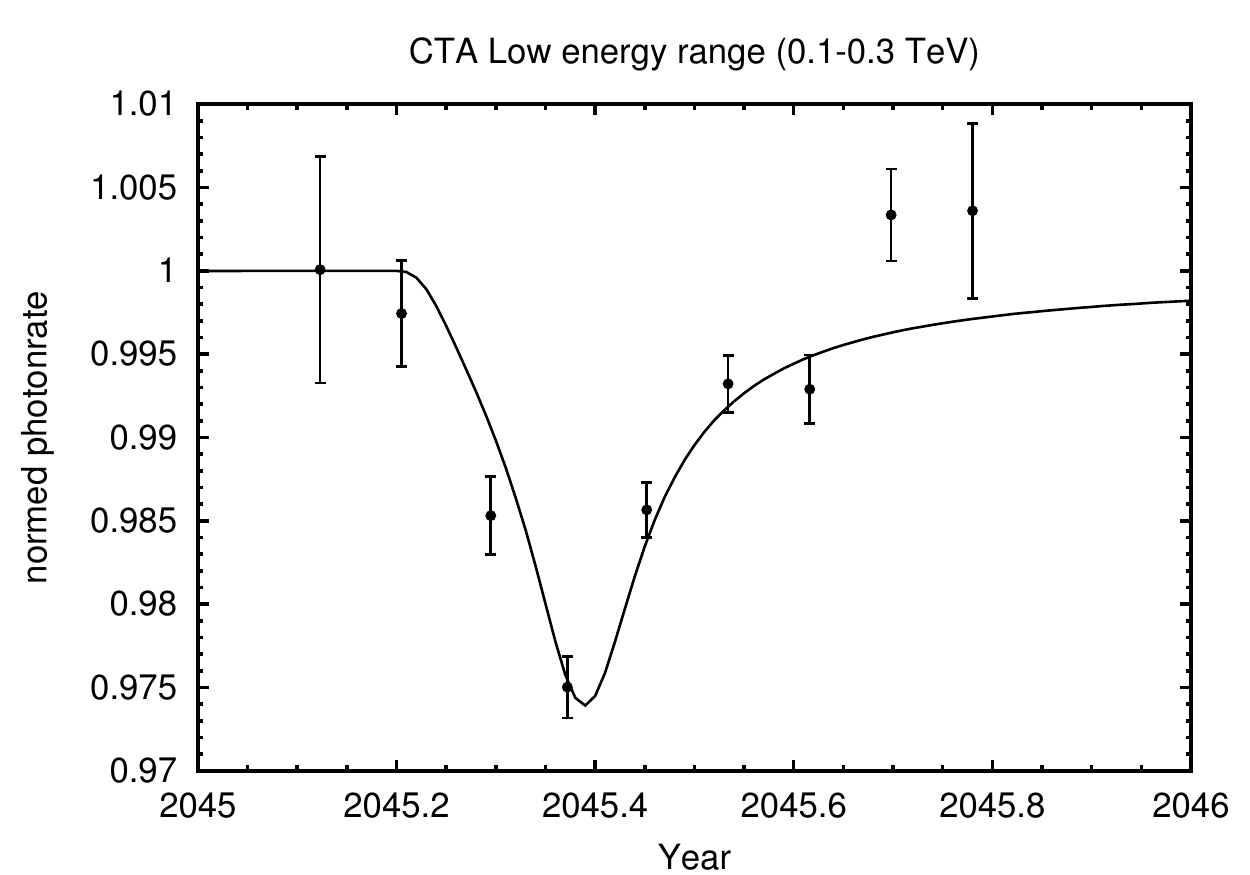} 
  \caption{Simulated Lightcurves: Low-energy band 0.1-0.3 TeV. The
    solid line is the calculated absorption for 0.2 TeV.}
  \label{fig:lightcurve_low}
\end{figure}

\section{Simulation of lightcurves}
\small
\begin{table}[b]
\begin{tabular}{lccc}
\hline
  & \tablehead{1}{c}{b}{H.E.S.S. }
  & \tablehead{1}{c}{b}{H.E.S.S. II}
  & \tablehead{1}{c}{b}{CTA}\\
& [Hz] &[Hz] &[Hz]\\
\hline

0.1-0.3 TeV \\energy band & $7.8\times 10^{-3}$  & $9.0\times 10^{-3}$ & $7.5\times 10^{-1}$\\
3-10 TeV\\ energy band & $6.5\times 10^{-4}$ & $2.1\times 10^{-3}$ & $1.8\times 10^{-2}$\\
\hline
\end{tabular}
\caption{Derived photon rates $R$ (without absorption)}
\label{tab:photon}
\end{table}
\normalsize
The flux modulation due to the absorption is very likely in the range
<10\% (based on the estimates of the underlying uncertainties) for a \gam origin in the SMBH. Current limits on
variability \cite{2006Natur.439..695A} from the GC are not sufficient
to put a meaningful constraint on the GC origin of the \gam
emission. In order to investigate the potential for discovering the
signature of pair eclipses, we simulate the measured lightcurve for
future ground based installations like H.E.S.S. II and CTA. For
completeness, we also simulated the expected H.E.S.S. lightcurve. The simulations shown here are for the star S14 which is the
candidate with the largest absorption effect. To simulate the
lightcurves we first derived the expected photon rate $R$ (equation
\eqref{eq:photon}) for two energy bands (low: $E_{1}=100 \, \text{GeV}$ to
$E_{2}=300 \, \text{GeV}$, high: $E_{1}=3 \, \text{TeV}$ to $E_{2}=10
\, \text{TeV}$).

\begin{equation}
R(t)=\int\limits_{E_{1}}^{E_{2}} dE \ A_{eff}(E)  \phi(E) \ \exp(-\tau(E,t))
\label{eq:photon}
\end{equation}

The parts of this equation are the following:

\begin{itemize}
\item
  $A_{eff}$: the effective area of the telescope:
  \begin{itemize}
  \item
    H.E.S.S. taken from \cite{2006A&A...457..899A}
  \item
    H.E.S.S. II taken from \cite{Punch2005}
  \item
    for CTA we assumed a 25-fold increase of the collection area
    compared to H.E.S.S.
  \end{itemize}
\item
  $\phi(E)$: the spectrum of the GC following a power law with index 2.25 \cite{2006PhRvL..97v1102A}
\item
  $\tau (E,t)$: taken from equation \eqref{eq:tau}
\end{itemize}

Table \ref{tab:photon} shows the derived photon rates for the three
IACT configurations. The values given there are without absorption. 
The derived values are then Gaussian distributed within the
error $\sqrt{R}$ (the expected number of photons in each energy bin is
sufficently large to justify Gaussian errors). Every lightcurve is integrated over one month
taking into account the theoretically possible observation time for the
Galactic Center (at the H.E.S.S. site in
Namibia). Figure \ref{fig:lightcurve_high} and
\ref{fig:lightcurve_low} show the result of the simulation for the three different
telescopes configurations. In the lower energy band the absorption has its maximum while in the
higher band almost no absorption is expected. The higher
band can be used to eliminate to first order all systematic
effects which could mimic variability.  For H.E.S.S. and H.E.S.S. II it is not possible to
determine the absorption feature within the errors. A large telescope array, like CTA, should be able to measure the effect.

\section{Conclusions}

We have shown that some S-stars in
the vicinity of \sgr   \ are getting close enough to the line of sight to
cause a notable absorption effect. The absorption leads to a dimming of the measured
\gam flux at energies between 100 GeV to 300 GeV. We presented
simulations of lightcurves for present and future IACTs. The current
generation of IACTs is not sensitive enough to detect
the effect, but the required sensitivity will be achieved with high
count-rate experiments ($\sim 1$ Hz) with large collection areas, like CTA.
If this time- and energy-dependent feature can be
measured in the spectrum of the TeV-emission from the GC, then it is
possible to limit the origin of the \rays  to the direct vicinity
of \sgr  (within approximately
1000 Schwarzschild radii).


\begin{theacknowledgments}
  Attila Abramowski would like to thank the Bundesministerium f\"ur Bildung
  und Forschung for making the participation to the conference
  possible.
\end{theacknowledgments}



\bibliographystyle{aipproc}   

\bibliography{references}


\end{document}